\documentclass[reprint,amsmath,amssymb,aps,prl,superscriptaddress,showpacs,citeautoscript,floatfix]{revtex4-1}
\usepackage{graphicx}
\usepackage{dcolumn}
\usepackage{bm}
\usepackage{amsmath}
\usepackage{amssymb}
\usepackage[]{units}
\usepackage{subcaption}
\usepackage{booktabs}
\begin{document}

\title[Real area of contact]{\textit{Ab-initio} calculation of the real contact area on the atomic scale}

\author{M Wolloch}
\email{mwo@cms.tuwien.ac.at}
\affiliation{Institute of Applied Physics, Vienna University of Technology, Gu\ss hausstra\ss e 25--25a, 1040 Vienna, Austria}
\affiliation{Austrian Center of Competence for Tribology, Viktor-Kaplan-Stra\ss e 2, 2700 Wiener Neustadt, Austria}
\author{G Feldbauer}
\affiliation{Institute of Applied Physics, Vienna University of Technology, Gu\ss hausstra\ss e 25--25a, 1040 Vienna, Austria}
\affiliation{Austrian Center of Competence for Tribology, Viktor-Kaplan-Stra\ss e 2, 2700 Wiener Neustadt, Austria}
\author{P Mohn}
\affiliation{Institute of Applied Physics, Vienna University of Technology, Gu\ss hausstra\ss e 25--25a, 1040 Vienna, Austria}
\author{J Redinger}
\affiliation{Institute of Applied Physics, Vienna University of Technology, Gu\ss hausstra\ss e 25--25a, 1040 Vienna, Austria}
\author{A Vernes}
\affiliation{Institute of Applied Physics, Vienna University of Technology, Gu\ss hausstra\ss e 25--25a, 1040 Vienna, Austria}
\affiliation{Austrian Center of Competence for Tribology, Viktor-Kaplan-Stra\ss e 2, 2700 Wiener Neustadt, Austria}

\begin{abstract}
We present a novel approach to determine the onset of contact between a tip and a surface. The real contact area depending on the distance is calculated using Bader's quantum theory of atoms in molecules. The jump to contact, which is often observed in atomic force microscopy experiments, is used as an indicator for the initial point of contact, which in turn is defined by atomic relaxations and thus without the need of external parameters. Within our approach the contact area is estimated by evaluating the zero flux surfaces between the touching Bader-atoms, where the necessary electronic density cutoff for the Bader-partitioning is calculated to depend on the initial point of contact. Our proposed approach is therefore completely \textit{ab-initio} and we are able to define and calculate the real area of contact without imposing restrictions or free parameters. As a prototype system we choose a tip made of a ten atom tungsten pyramid above a moir\'{e} layer of graphene on an fcc iridium (111) substrate. We find that the contact area depends exponentially on the effective distance between the tip apex and the surface atom directly below within the atomically relaxed nanosystem.
\end{abstract}


\maketitle

\section{Introduction}
\label{sec:Int}

Historically the introduction of the concept of a real contact area in
1939 by Bowden and Tabor, which is substantially smaller than the
nominal one, was a huge step forward in understanding the laws of
friction~\cite{bowden:39}. Since then a multitude of methods have been
used to predict the pressure dependent real contact area for rough
surfaces.

In 1881 Hertz laid the foundations of contact mechanics by describing the junctions between non-adhesive, homogeneous elastic solids of simple shapes~\cite{hertz:1881}. Nearly a hundred years later Johnson,
Kendall, and Roberts (JKR) included short-range adhesion forces
inside the contact which lead to larger contact areas compared to
Hertz's model~\cite{johnson:71}. In contrast, Derjaguin,
Muller and Toporov (DMT) assumed that the Hertzian contact area
remains undeformed while long-range adhesion forces are acting outside
the contact zone~\cite{derjaguin:75}. Tabor was able to show that both the JKR and the DMT model can be viewed as limiting cases of a more general model, with JKR suitable for soft materials with large adhesion and DMT for hard materials with low adhesion~\cite{tabor:77}. Additional work on this unification was done by Maugis using Dugdale potentials~\cite{maugis:92}. The analytical solution by Maugis, however, produces rather cumbersome equations, which can be approximated with high accuracy by the generalized transition equation derived by Carpick, Ogletree and Salmeron~\cite{carpick:99}. Their result also describes the transition between
the DMT and the JKR models, but with simpler expressions which can be
applied more straightforwardly to experimental data and only differs
from the Maugis-Dugdale model within the unstable low load
region. Generally these theories agree that the contact area $A$ between a single sphere and a flat surface is a sublinear function of the load $L$, see e.g.~the original Hertz prediction of $A(L)\sim L^{2\,/\,3}$. These continuum
mechanics models are undoubtedly very successful and play an important
role in both theoretical and experimental work in tribology. However, on the atomic scale, as tested for example in AFM experiments, the size of the the contact approaches the size of the involved atoms and thus models based on continuum mechanics are hardly applicable~\cite{luan:05}. The apparent success of the Maugis-Dugdale model, which is still widely used to interpret atomic scale AFM experiments, can be attributed to its flexibility provided by three fitting parameters~\cite{dong:13}. Hence, atomistic methods are of great value in assessing the validity of the results and provide a much needed tool to increase the understanding of the real contact area.

In recent studies and discussions a need for characterizing the contact area on an atom-by-atom basis was expressed and various strategies to estimate the number of atoms in contact were proposed \cite{luan:05,mo:09,cheng:10,eder:11,eder:14}. However, it is not trivial to decide at which distance two atoms are in contact or how big the resulting contact area should be. One possibility is to define a certain inter-atomic distance $a_0$ below which contact should be established, however, this just shifts the problem to find the correct distance $a_0$. A common method is to identify $a_0$, and thus the onset of contact, as the beginning of repulsion between the two observed atoms~\cite{burnham:91,cheng:10}. To this end, classical molecular dynamics (MD) studies using Lennard-Jones potentials are often employed, sometimes without the attractive part if the surfaces of interest are non-adhesive. In our view, this method is not ideal for describing the onset of contact on the nanoscale for the following two reasons: (i) if only repulsive interaction is a sign of contact, the atoms in a solid or molecule at \unit[0]{K} are not in contact with each other; (ii) in AFM experiments one often observes a ``jump to contact'', where either the tip, or some part of the surface below or both jump towards each other because of strong attractive interactions. If now the tip support is lowered further, the distance between the surface and the tip apex might get even smaller as the chemical binding becomes stronger. It seems unreasonable to argue that the tip is not in contact with the surface after the jump, just because the interaction is still attractive.

In a study from 2009, Mo, Turner, and Szlufarska examined the contact area between hydrogenated amorphous carbon tips of up to \unit[30]{nm} radius and a flat hydrogenated diamond surface~\cite{mo:09}. For this large scale finite temperature (\unit[300]{K}) MD study they used a reactive empirical bond-order (REBO) potential~\cite{brenner:02} to model the chemical forces and added an analytical switching function to include van der Waals (vdW) like long-range forces. The multi-asperity picture of nanoscale contact presented in their publication relies on the assumption that contact is established between the atoms that are interacting chemically through the REBO potential while a much larger part of the tip is attracted to the surface via the vdW forces. These ideas are further discussed in a follow-up publication by Mo and Szlufarska~\citep{mo:10}. Now an atomic contact area $A_{\text{at}}$ is attributed to
every chemically interacting atom of the tip leading to the total real
contact area $A_{\text{real}}=N_{\text{at}} A_{\text{at}}$, with
$N_{\text{at}}$ being the number of involved atoms. Due to atomic scale roughness in the amorphous tip this real contact area may be significantly smaller than the expected contact area of a smooth asperity, $A_{\text{asp}}$, which is defined as the envelope over the contact points.
Concluding, Mo, Turner and, Szlufarska pointed out how
important atomic corrugations are for an accurate estimation of the
real contact area, and they underlined the need for accurate
computational approaches. Nevertheless, a few points remain to be analyzed further. In a realistic, continuous potential it might be hard to distinguish between long-range dispersion forces and chemical forces, thus making it difficult to define the number of atoms in contact. Furthermore, the inherent assumption that all contributions from single atomic contacts $A_{\text{at}}$ are of equal size might not hold in systems with varying local load.

In the following, we will introduce an \textit{ab-initio} approach to
estimate the distance-dependent real contact area between a tip and a
surface. To prove the feasibility of our approach, we choose a
realistic and thus rather complex system, which has previously been
used to explain contrast inversion between constant current scanning
tunneling microscope and constant frequency AFM images of graphene
moir\'e on metals~\cite{voloshina:13,garhofer:phd}. We employ van der
Waals corrected density functional calculations and use
Bader-partitioning of the electronic charge density to identify
accurate atomic volumes and surfaces in different chemical
surroundings.

\section{Computational methods}
\label{sec:Comp}
\begin{figure}[htbp]
	\centering
        \begin{subfigure}[b]{0.75\linewidth}
                \includegraphics[width=\textwidth]{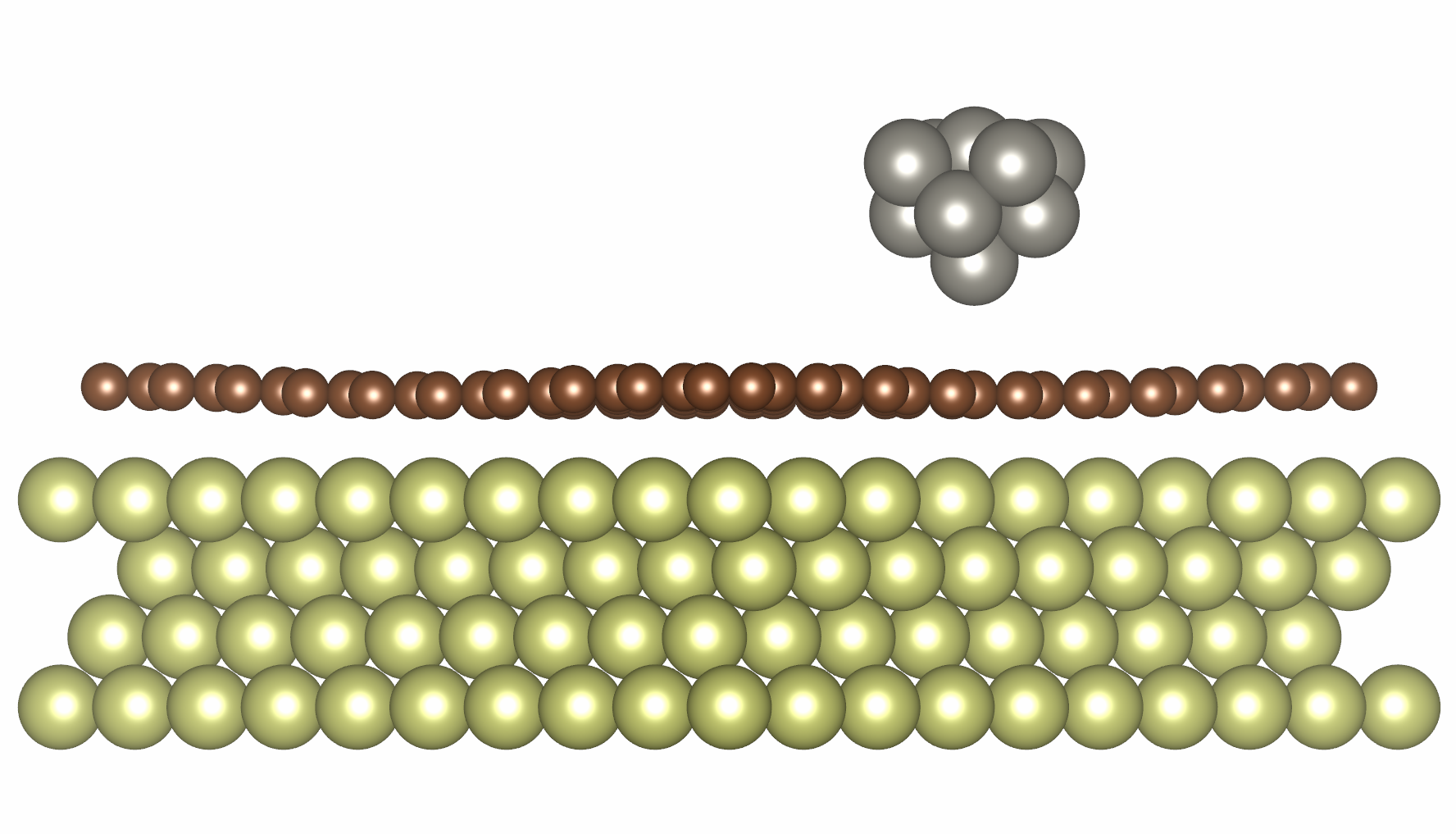}
                \caption{Side view}
                \label{fig:IrGrW_unitCell_a}
        \end{subfigure}        
        
        \begin{subfigure}[b]{0.75\linewidth}
                \includegraphics[width=\textwidth]{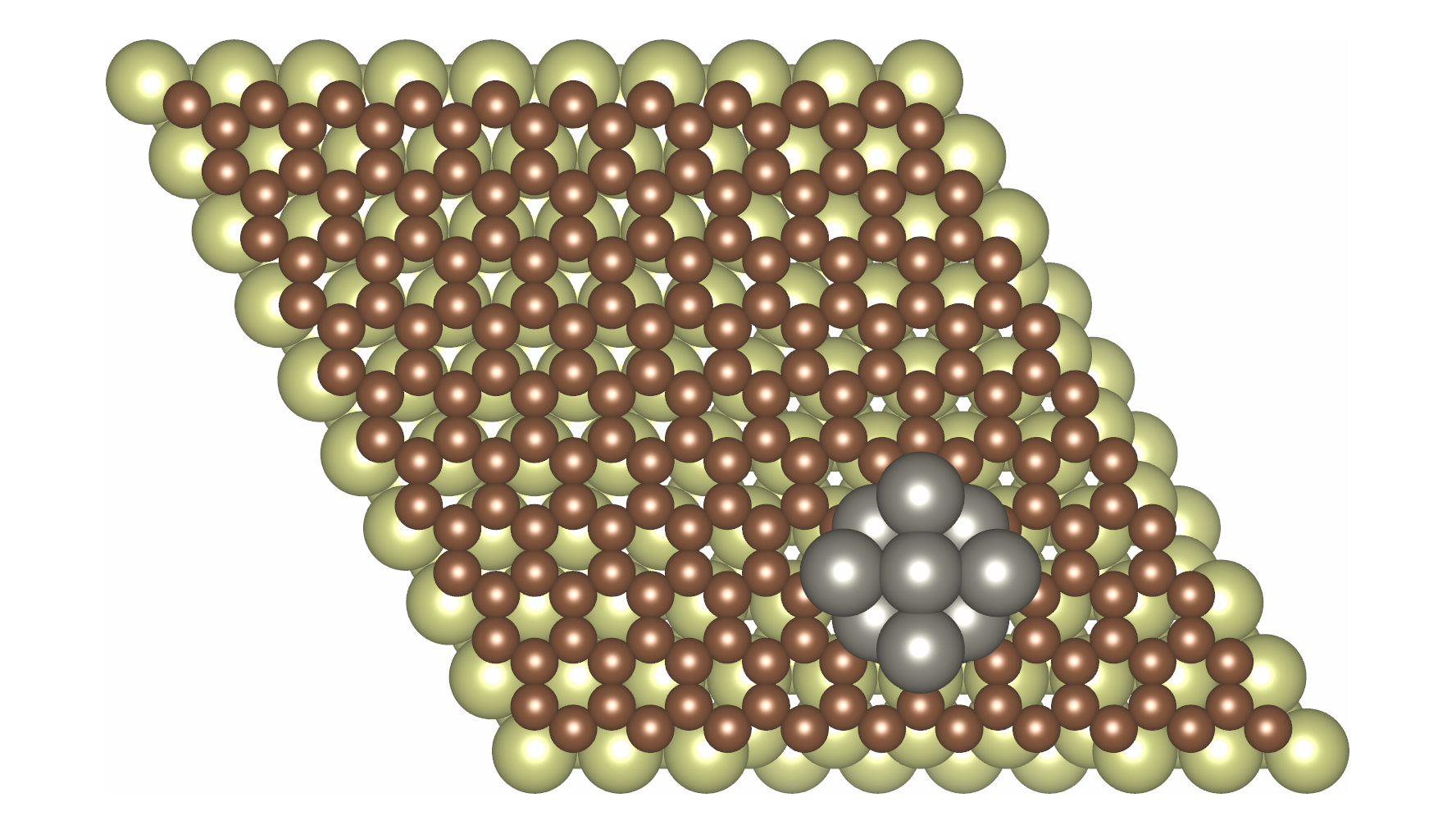}
                \caption{Top view}
                \label{fig:IrGrW_unitCell_b}
        \end{subfigure}
    \caption{Side (a) and top (b) view of a tungsten tip on
      graphene/Ir(111). Iridium atoms are shown in yellow,
      carbon in brown and tungsten in grey.}
	\label{fig:IrGrW_unitCell}
\end{figure}

The simulation cell consists of 4 layers of a \( 9 \times 9 \) iridium(111) substrate covered by a \( 10 \times 10 \) graphene layer forming a moir\'e structure with an average separation of \unit[3.42]{\AA} and a corrugation of \unit[0.35]{\AA}. The lattice constants obtained with
the optB86b-vdW functional~\cite{klimes:10,klimes:11}, $a_{\mathrm{Ir}}=\unit[2.735]{\AA}$ and $a_{\mathrm{Gr}}=\unit[2.465]{\AA}$, are close to the experimental values of \unit[2.71]{\AA} and \unit[2.46]{\AA}, respectively. The mismatch of the structure is very small at $10 a_{\mathrm{Gr}} - 9 a_{\mathrm{Ir}} = \unit[0.015]{\AA}$~\cite{garhofer:phd}. The tungsten AFM tip was modeled as a ten atom pyramid with one atom at the apex, four in the next layer and five at the top. The contact site studied was an on-top position in a top-hcp region of the moir\'e structure. This means that the tip apex atom is positioned directly vertical over a carbon atom in a region where each carbon atom is either directly over an iridium atom or over an iridium hcp position. The simulation cell, containing 534 atoms, is shown in figure~\ref{fig:IrGrW_unitCell}. Relaxations were allowed for the graphene layer and the bottom five atoms of the tip, keeping the iridium substrate and the top layer of the tip rigid at their initial relaxed positions. Relaxations of the iridium substrate during movement of the tip have been neglected due to the small binding energy ($\sim$\unit[80]{meV} per carbon atom) of the mainly physisorbed graphene layer, which makes any effect of the iridium substrate on the relaxations of the graphene unlikely.  The topmost layer of the tungsten pyramid, on the other hand, needs to be held fixed to control the distance between the tip and the graphene sheet.

All calculations were performed within density functional theory (DFT) employing the Vienna Ab-Initio Simulation Package \textit{VASP}~\cite{kresse1993,kresse1994a,kresse1996a,kresse1996b} using the Projector Augmented-Wave (PAW) method~\cite{bloechl1994,kresse:98}. To include van der Waals (vdW) forces, which are relevant in this system, the optB86b-vdW functional was employed~\cite{klimes:10,klimes:11}. This vdW density functional has been applied to a wide range of materials and proven to be of good accuracy~\cite{chakarova:06, sony:07,carrasco:11, mittendorfer:11, graziano:12,bedolla:14a, bedolla:14b}. The Brillouin zone sampling was performed on a \(\Gamma\)-centered \(3 \times 3 \times1 \) k-grid, with a smearing of \unit[0.1]{eV} using the method of Methfessel and Paxton to first order~\cite{methfessel:89}. To ensure good accuracy for the Bader-partitioning scheme, the electronic charge density was calculated on a dense mesh of $432 \times 432 \times 448$ points in the simulation cell. Electronic energies were converged to $\unit[10^{-6}]{eV}$ and the ionic relaxations were stopped after converging forces between the interacting atoms to better than \unit[0.01]{eV/\AA}. Following the investigation by Garhofer, who also provided us with initial structural data~\cite{garhofer:phd}, we choose a plane wave cutoff of \unit[300]{eV}, which is the minimum recommended value for carbon, and at the same time larger than the suggested value for iridium and tungsten.

To partition the electronic charge density $\rho$ in our simulation cell into single atoms we use Bader's quantum theory of atoms in molecules (QTAIM)~\cite{BOOK_bader:90}. In contrast to other similar
approaches~\cite{hirshfeld:77,bultinck:07,bickelhaupt:96,fonseca:04,mayer:04,becke:88},
Bader's method produces non-overlapping atomic domains with
well-defined boundaries, which are perfectly suited to analyze the
contact between two adjacent bodies. The necessary and sufficient condition that needs to be fulfilled to define the boundaries of a selected atom according to the QTAIM is formulated using the basic quantity in DFT, namely the electronic charge density $\rho$ and is given as~\cite{bader:85},
\begin{equation}
\nabla \rho({\bf r})  \cdot {\bf n}({\bf r})= 0 \quad \forall {\bf r}\in S({\bf r}) \quad.
\label{eq:bader1}
\end{equation}
Here $S$ is the boundary surface of the atom and ${\bf n}(\bf r)$ is the unit vector normal to this surface. The condition states that the flux of the gradient field of the charge density, $\nabla \rho$, through the boundary surface $S$ must vanish, which is thus called a zero flux surface. The Bader
analysis in this work was performed with the code developed by
Henkelman, Sanville, and Tang which is directly compatible with the
format of \textit{VASP}-output
files~\cite{henkelman:06,sanville:07,tang:09}.

\section{Results}
\label{sec:Results}

If one seeks to define the real contact area on an atomic scale, it is quite natural to think about the size, shape, and deformations of the involved atoms. Once the size and shape of all atoms in the contact region are determined, the calculation of the real contact area is reduced to a simple summation of the regions that are in contact, provided one can distinguish unambiguously between the two contacting bodies. Since $\rho$ formally is non-zero everywhere, the Bader-atoms at the surfaces of the contacting bodies extend into the vacuum region to infinity or until they encounter another atom. This would mean that contact between two bodies is established at all distances, which is a clearly unphysical result, unless one defines a density cutoff. This density cutoff $\rho_{\mathrm{cut}}$ cannot be chosen arbitrarily, since it directly influences the contact area.

\begin{figure}[htbp]
\centering
\includegraphics[angle=0, width=0.65\linewidth]{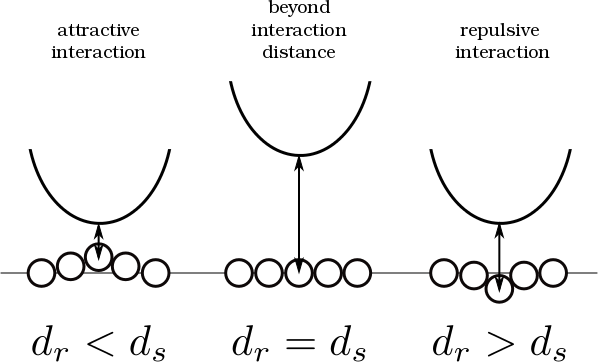}
\caption{Sketch of possible tip-surface interactions. If the tip is far away from the surface (middle panel), atomic relaxations will not have an effect on the distance between the tip and the surface. If the tip gets closer, the surface will interact with the tip and will either jump towards the tip, if the interaction is attractive (left panel), or will get depressed, if the interaction is repulsive (right panel). This effect can be used to define the onset of contact.}
 \label{fig:tips}
\end{figure}

A possibility to extract a value for $\rho_{\mathrm{cut}}$ is to analyze the interaction potential between the tip and the surface, divide it into a long- and a short-range part and define the contact at the onset of the short-range interaction, analogous to Mo, Turner, and Szlufarska~\cite{mo:09}. This procedure defines a density cutoff $\rho_{\mathrm{cut}}$ so that the Bader-partitioning yields contact only after the onset of short-range interactions. However, as long-range interactions are included
implicitly in the exchange-correlation potential that we use (see
section Computational Methods), the separation into a long- and a
short-range part is not straightforward.
A more unambiguous way to define the onset of contact is to analyze the atomic relaxations that happen if the tip is lowered towards the surface. We distinguish the distance for the static, unrelaxed system ($d_s$) and the relaxed distance ($d_r$), which are both measured between the tip apex atom and the carbon atom directly beneath it. For large distances
no relaxations will happen, although there might be attraction due to
vdW forces, and the distance $d_r$ in the relaxed system will be equal
to the (static) distance $d_s$ measured before relaxing the system. At some point during the approach of the tip stronger forces will cause relaxations, which will result either in a ``snap'' or ``jump'' to contact (a phenomenon often observed in AFM experiments) if the interaction is attractive, or in a depression of the surface layer if the interaction is purely repulsive. A sketch of this process is given in figure~\ref{fig:tips}. In any case, the $d_r$ versus $d_s$ curve will have a discontinuity at some distinct distance where the system begins to strongly interact. Below this distance the system will try to hold the ideal distance between tip and surface. It is straightforward to identify this discontinuity as the onset of contact.

\begin{figure}[htbp]
	\centering
        \begin{subfigure}[b]{0.85\linewidth}
                \includegraphics[width=\textwidth]{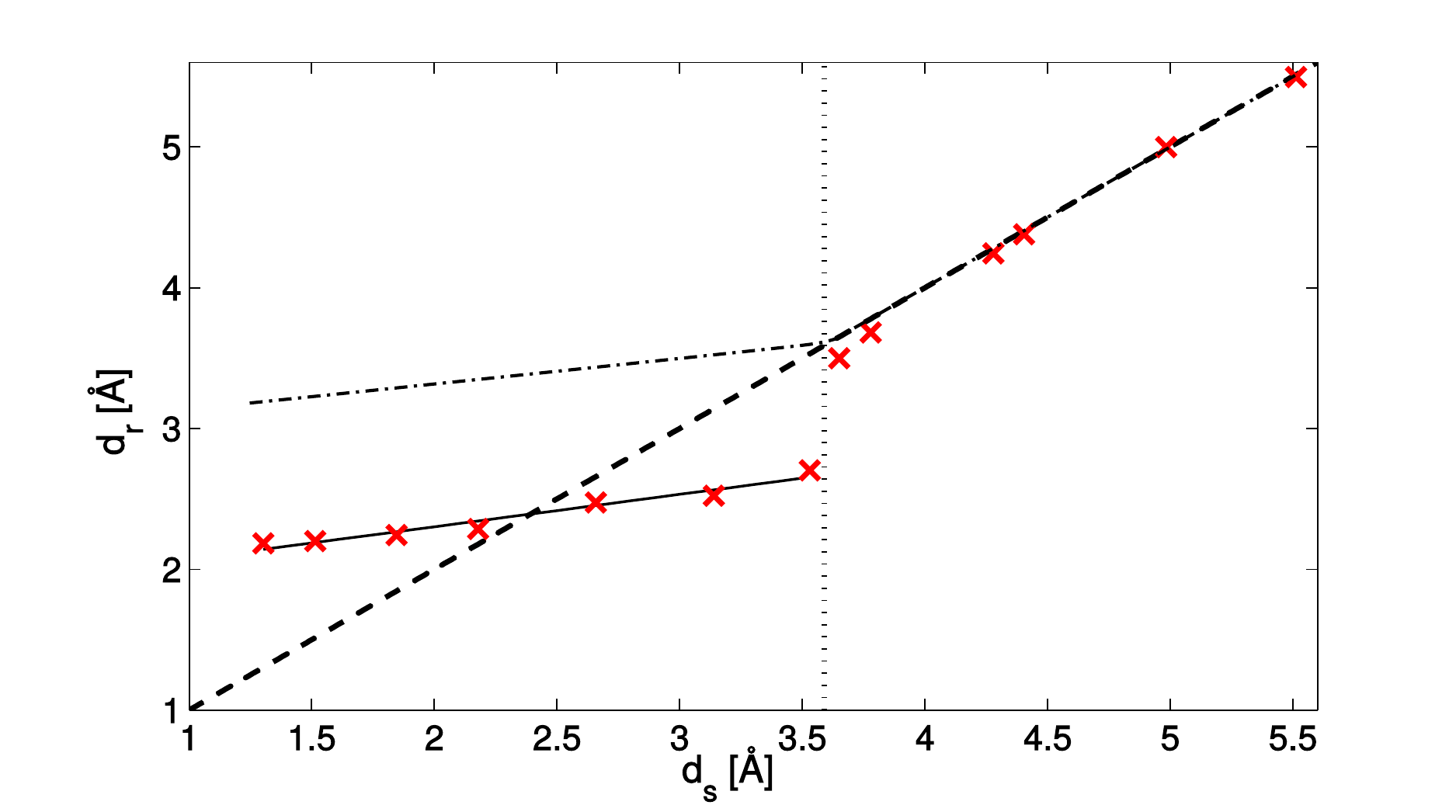}
                \caption{}
                \label{fig:jump}
        \end{subfigure}        
        
        \begin{subfigure}[b]{0.85\linewidth}
                \includegraphics[width=\textwidth]{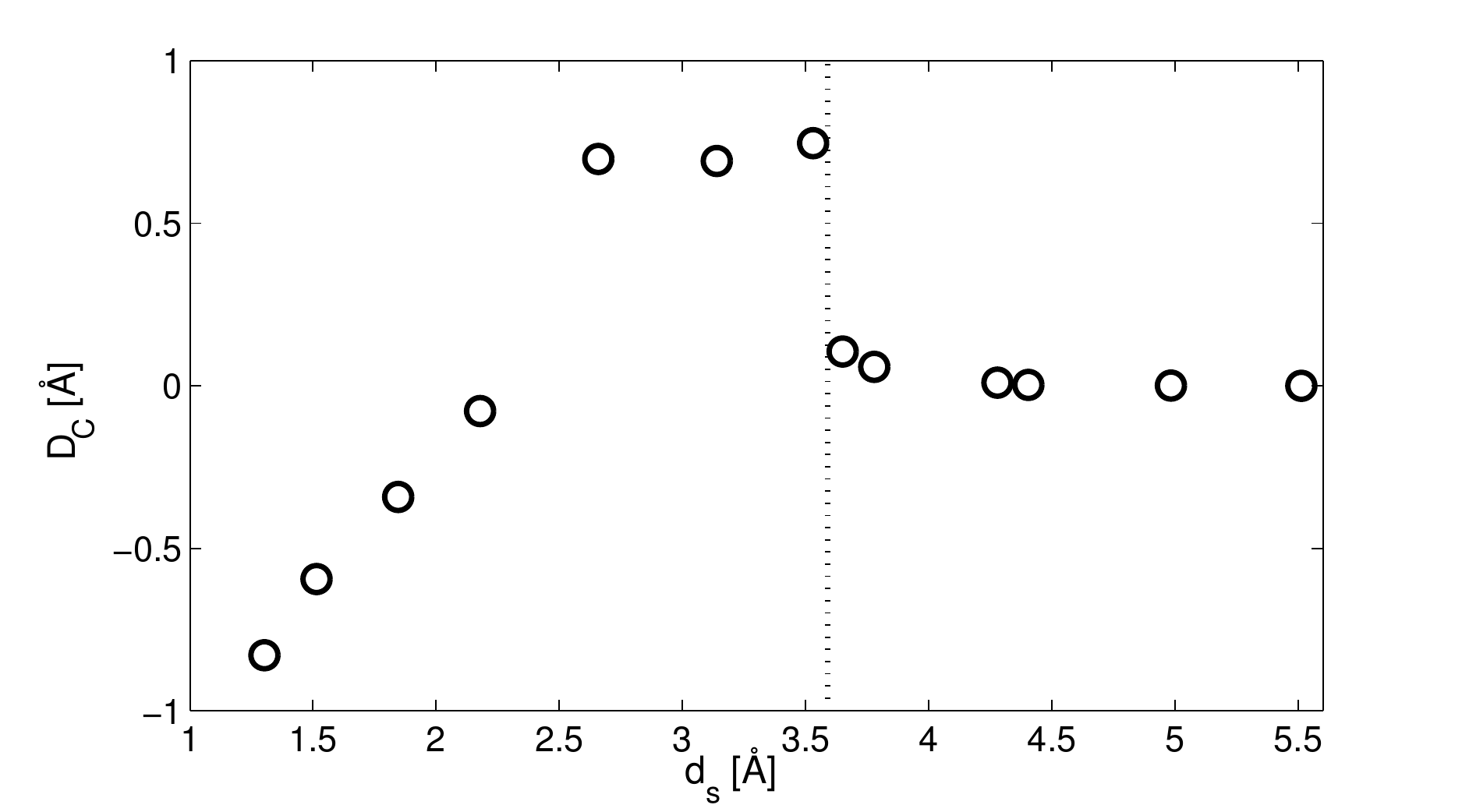}
                \caption{}
                \label{fig:c_pos}
        \end{subfigure}
\caption{(a) Distance $d_r$ in the relaxed system (red crosses) versus
  the (static) distance $d_s$ between a rigid tungsten tip approaching graphene/Ir(111). The dashed line gives $d_r=d_s$. Attractive interactions cause a jump to contact between  $d_s=\unit[3.65]{\AA}$ and $d_s=\unit[3.53]{\AA}$, which is marked with a vertical dotted line. After the jump to contact the relation between $d_r$ and $d_s$ is also linear (solid black line). This line crosses the $d_r=d_s$ line at the equilibrium point, where the graphene layer has relaxed back into its original shape. A hypothetical curve for a purely repulsive interaction is sketched by the dashed-dotted line to illustrate that the method is also viable if no jump to contact is occurring in the system. (b) Displacement $D_C$ of the carbon atom situated directly below the tip apex with respect to its initial position, versus the distance $d_s$. The jump to contact is again marked by a vertical dotted line.}
 \label{fig:jump+Cpos}
\end{figure}

In the examined system the interaction between the tip and the surface is attractive at the onset of contact and a jump to contact occurs between $d_s=\unit[3.65]{\AA}$ and $d_s=\unit[3.53]{\AA}$ (see figure~\ref{fig:jump}), such that $d_r$ is changing from \unit[3.5]{\AA} to \unit[2.7]{\AA}, accordingly. Most of the movement is done by the surface, which jumps upwards to meet the tip (see figure~\ref{fig:c_pos}). This means that when the tip support is lowered by only \unit[0.12]{\AA}, the distance between the tip apex and the surface is reduced by \unit[0.8]{\AA}. This allows us to define the onset of contact at $d_s\simeq \unit[3.6]{\AA}$ and to tune the value of the density cutoff $\rho_{\mathrm{cut}}$ accordingly. Note that an infinitely stiff cantilever is assumed in our calculations, as the uppermost atoms of our tip are kept rigid. In principle, the influence of a more compliant AFM apparatus on the initial jump to contact, however, could also be modeled by using a multiscale approach.

Once $\rho_{\mathrm{cut}}$ is selected the determination of the real contact area for each distance is straightforward. Since the partitioning code produces only Bader volumes rather than zero flux surfaces, we have to construct the contact area from these data. To this end the Bader-volumes of both contacting bodies are added up and by pairwise comparison of the respective values for neighboring grid points a point cloud forming the contact area is generated. The contact area can now be obtained by triangulation.

We calculated contact areas
for cutoff densities from \unit[10$^{-3}$]{e/\AA$^3$}, which is the
default cutoff in the partitioning code by Henkelman, Sanville and
Tang~\cite{henkelman:06,sanville:07,tang:09}, up to a cutoff of
\unit[10$^{-1}$]{e/\AA$^3$}. While the default value of $\rho_{\mathrm{cut}}=\unit[10^{-3}]{e/\AA^3}$ and other low cutoff densities are giving sizable contact areas for all distances, we approach the desired effect of establishing contact only for $d_s<\unit[3.6]{\AA}$, for a value of $\rho_{\mathrm{cut}}\sim\unit[10^{-2}]{e/\AA^3}$. Obviously, a very high $\rho_{\mathrm{cut}}$ is unphysical, as the number of electrons that are ``lost'' into the vacuum region increases with rising cutoff. This means that we want to select a value that is high enough to guarantee that the contact area $A$ is only non-zero after the snap to contact has occurred, but is otherwise as low as possible. We analyzed several values of $\rho_{\mathrm{cut}}$ ranging from \unit[$7.5\times 10^{-2}$]{e/\AA$^3$} to \unit[ $1.0 \times 10^{-2}$]{e/\AA$^3$} in order to find the lowest value that still satisfies these conditions, resulting in an optimal value of $\rho_{\mathrm{cut}}=5 \times 10^{-2}$ electrons per \AA$^3$.

Of course, changing the contact site of the tip away from a position directly above a carbon atom or to another section of the moir\'e structure could conceivably change the exact point of the jump to contact and thus modify the value of $\rho_{\mathrm{cut}}$. However, the obtained cutoff density of $\rho_{\mathrm{cut}}=5 \times 10^{-2}$ is low enough to result in an essentially flat graphene surface and hence changing the tip position should only marginally alter the computed contact area. As this paper is mainly concerned with the presentation of a new approach in defining and calculating the real contact area \textit{ab-initio}, and given the rather large computational effort \footnote{Calculations where carried out on the Vienna Scientific Cluster, typically using 16 computer nodes interconnected by an QDR InfiniBand network, each equipped with two 8-core AMD Opteron 6132HE \unit[2.2]{GHz} CPUs and \unit[32]{GB} of ECC DDR3 RAM. Computation times on this cluster, including pre-relaxation with a GGA functional, full relaxations with the optB86b vdW functional and final static calculations with the finer mesh to generate proper charge densities for the Bader partitioning, took up to 100 hours or 25600 corehours for each tip position.}, we decided against repeating our calculations on different contact sites and calculating an average.

In a preliminary calculation of the same tip on an fcc copper (111) surface we found an optimal charge density cutoff value of $5.3 \times 10^{-2}$ electrons per \AA$^3$, which is approximately the same as for the graphene/Ir(111) system.

As the optimized cutoff of \unit[$5 \times 10^{-2}$]{e/\AA$^3$} is 50 times larger than the default value it is important to check if it is still a reasonable number and does not give any unphysical results. We therefore calculated the nominal number of electrons that are assigned to the vacuum region and thus are not part of any Bader-atom. For the default $\rho_{\mathrm{cut}}$ of \unit[$1 \times 10^{-3}$]{e/\AA$^3$} only about half of an electron is not represented by a Bader-atom. For the cutoff value needed for the calculation of the contact area, \unit[$5 \times 10^{-2}$]{e/\AA$^3$}, this number is increased to nearly 30 electrons. Although this value seems to be very large, one has to consider the total system size, which includes 3776 electrons. Thus, the relative number of ``missing'' electrons is below 0.8\%. We also evaluated the Bader-radii $R_\mathrm{B}$ of a single tungsten atom and a single carbon atom in a box. The value for tungsten of \unit[2.9]{\AA} obtained for $\rho_{\mathrm{cut}}=5 \times 10^{-2}$ electrons per \AA$^3$ is more than twice as large as the empirical atomic radius, \unit[1.35]{\AA}~\cite{slater:64}, and about \unit[1]{\AA} larger than the calculated atomic radius of \unit[1.93]{\AA}~\cite{clementi:67}. Reported values for the atomic radius of carbon reach from the calculated value of \unit[0.67]{\AA}~\cite{clementi:67}, over the empirical value of \unit[0.70]{\AA}~\cite{slater:64} to a van der Waals radius of \unit[1.70]{\AA}~\cite{bondi:64}. As for tungsten, also the carbon Bader-atom radius for a cutoff density of $5 \times 10^{-2}$ electrons per \AA$^3$ is significantly larger than all these reported values at \unit[2.30]{\AA}. This ensures that neither the tip atoms nor the surface Bader-atoms are artificially small. It also shows that it is questionable to assume that contact between two bodies is established only after the surface atoms overlap if one uses spherical atoms and traditional radii.

It is worthwhile to compare our approach to the onset of contact with the method by Mo, Turner, and Szlufarska~\cite{mo:09}, which uses the beginning of short-range interaction as a criterion for contact. As already mentioned, the distinction between long-range and short-range forces is not trivial, but it might be approximated by disabling the long-range contributions in the correlation potential of the optB86b-vdW functional. We calculated the corresponding energies at the vdW relaxed positions and fitted the data with a Morse function~\cite{morse:29}. The interaction strength of this short-range potential at the jump to contact is 2.0\% of the total potential depth which could be classified as the ``beginning of the interaction''. This means that our approach, at least for the system investigated here, is in accordance with the approach of reference~\citenum{mo:09}. However, an interaction strength of 1\%, 5\% or even 10\% of the short-range binding energy could also be reasonably selected as the ``beginning of the interaction'', each leading to different results. This highlights the advantages of using the jump to contact as the criterion for the initial point of contact, as no further assumptions are needed proceeding in this manner.

\begin{figure}[htbp]
\centering
\includegraphics[angle=0, width=1.0\linewidth]{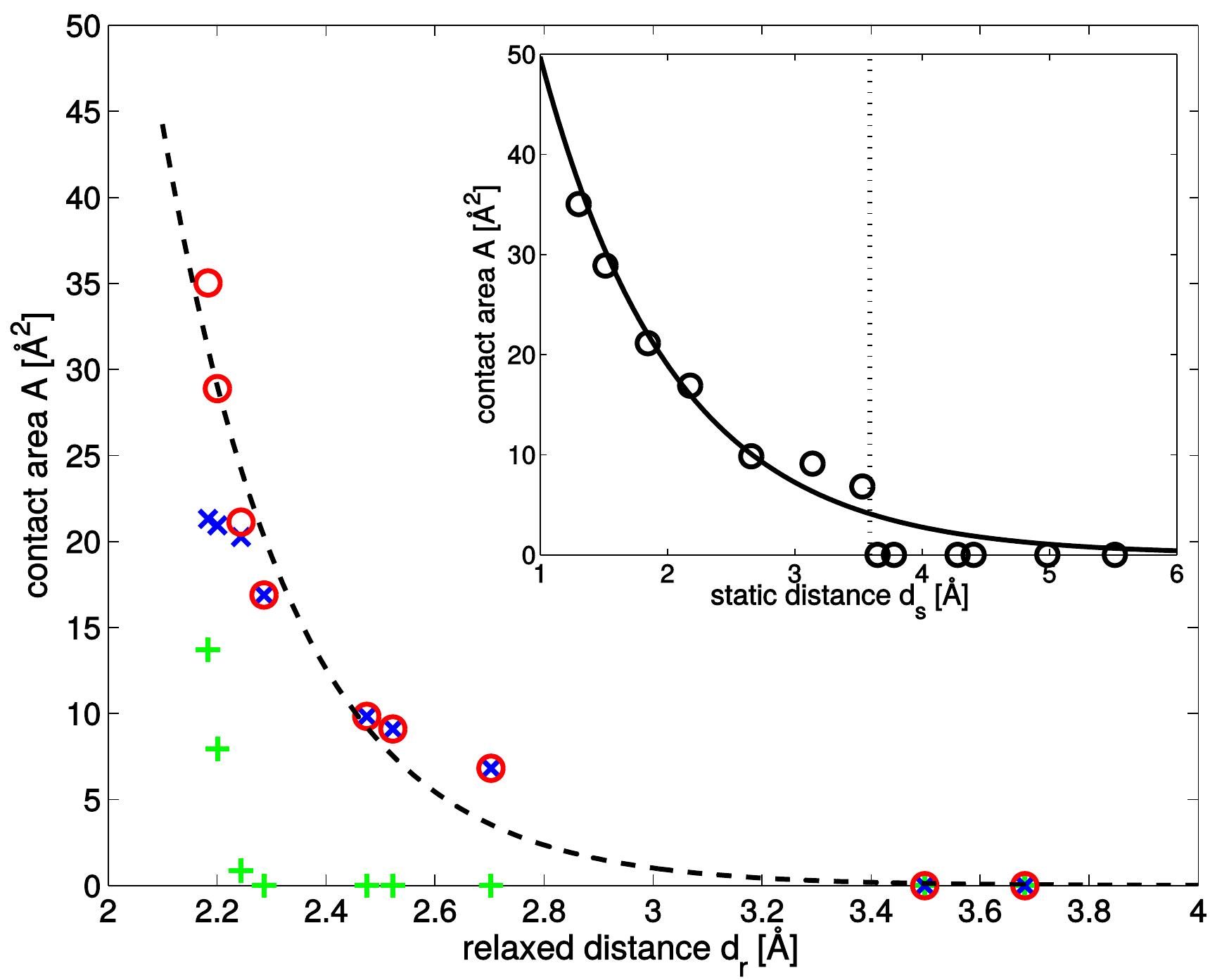}
\caption{{\it Ab-initio} real contact area $A$ obtained for
  $\rho_{\mathrm{cut}} = 5\times 10^{-2}$ e/\AA$^3$ versus the
  distance $d_r$ in the relaxed system between the tungsten tip and
  graphene/Ir(111).  Blue crosses give the contribution of the tip
  from the apex atom and green plus signs show the contribution from
  the second layer (four atoms). The total contact area (red circles)
  is given by the sum of these two contributions and is fitted by an
  exponential (dashed line). The inset shows the total contact area versus the static distance $d_s$. The solid line is an exponential function resulting from equation (\ref{eq:Ar}) and the linear relation between $d_r$ and $d_s$ (see figure~\ref{fig:jump}). The dotted vertical line marks the jump to contact.}
 \label{fig:IrGrW_A_vs_dr}
\end{figure}

Figure~\ref{fig:IrGrW_A_vs_dr} shows a decomposition of the contact area into contributions of the tip apex (one atom) and contributions from the second tip layer (four atoms). We find that the second layer only contributes to the total contact area for the three closest distances but is then responsible for nearly all of the increase. The dashed line in figure~\ref{fig:IrGrW_A_vs_dr} is an exponential fit of the form
\begin{equation}
A(d_r)=A_\Delta e^{ -\lambda_r \left( d_r-\Delta_r \right)} \quad ,
\label{eq:Ar}
\end{equation}
to the 7 non-zero data points (red circles) with the coefficients $\lambda_r \simeq \unit[4.2]{\AA^{-1}}$ and $\Delta_r \simeq \unit[3.0]{\AA}$. The factor $A_\Delta=\unit[1]{\AA^2}$ is included for dimensional reasons and has not been used as a fitting parameter. Although the exponential fit is not perfect, the agreement with the data is certainly reasonable, especially considering that only two fitting parameters were used. Also the point of vanishing contact is predicted well, although only points of positive contact area were considered for the fit.
As there is a linear relation between $d_r$ and $d_s$ in the region where contact is established ($d_r=\kappa d_s + \delta=0.23 d_s + 1.84$; see solid black line in figure~\ref{fig:jump}), it is also possible to express the contact area $A$ through the static distance $d_s$, which is easier accessible in experiments through the vertical displacement of the tip support. In the relation $A(d_s)=A_\Delta\exp\left[-\lambda_s \left( d_s-\Delta_s\right)\right]$, the decay constant is smaller than in $A(d_r)$ with $\lambda_s=\lambda_r \kappa\simeq \unit[1.0]{\AA^{-1}}$, while $\Delta_s=(\Delta_r - \delta) / \kappa \simeq \unit[5.0]{\AA}$ is increased compared to $\Delta_r$, and $A_\Delta=\unit[1]{\AA^2}$ is the same dimensionality factor as before. This relation is plotted in the inset of figure~\ref{fig:IrGrW_A_vs_dr}. Quite naturally only the region after the jump to contact is represented well.

\begin{figure*}[htbp]
	\centering
        \begin{subfigure}[b]{0.4\linewidth}
                \includegraphics[width=\textwidth]{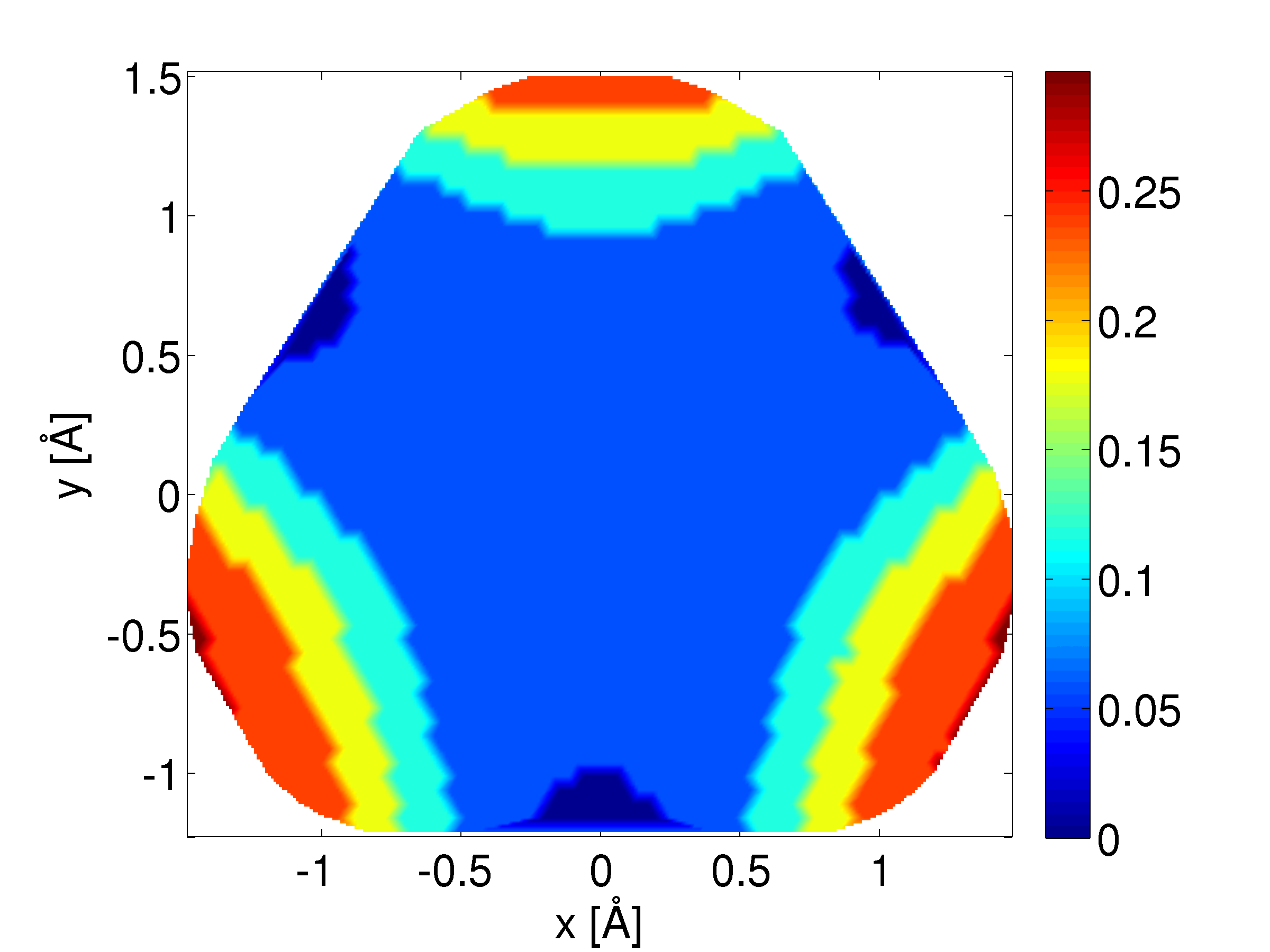}
                \caption{$d_s=\unit[3.53]{\AA}$}
                \label{fig:IrGrW_areas_1}
        \end{subfigure}
        \begin{subfigure}[b]{0.4\linewidth}
                \includegraphics[width=\textwidth]{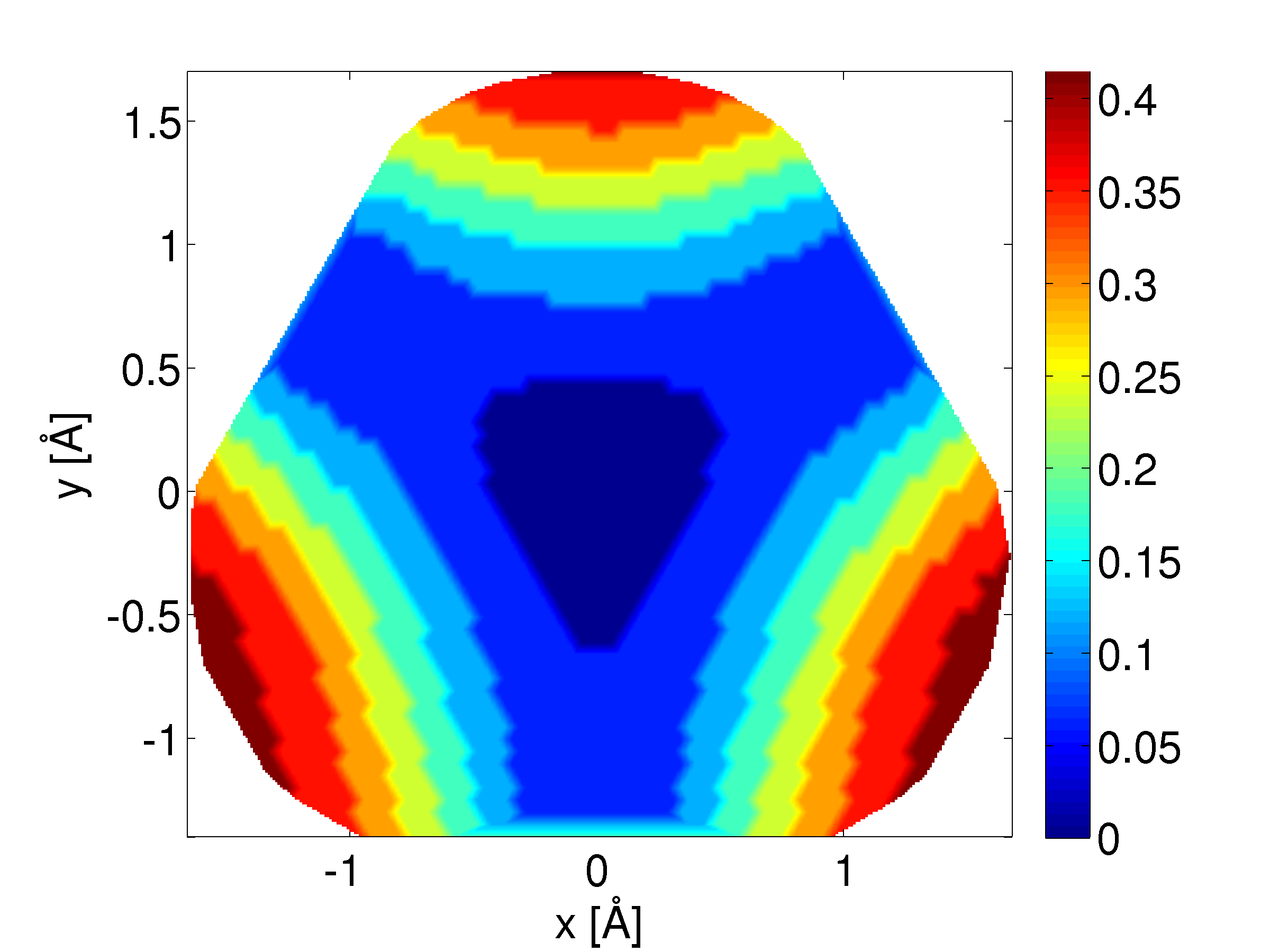}
                \caption{$d_s=\unit[3.14]{\AA}$}
                \label{fig:IrGrW_areas_2}
        \end{subfigure}        
        
        \begin{subfigure}[b]{0.4\linewidth}
                \includegraphics[width=\textwidth]{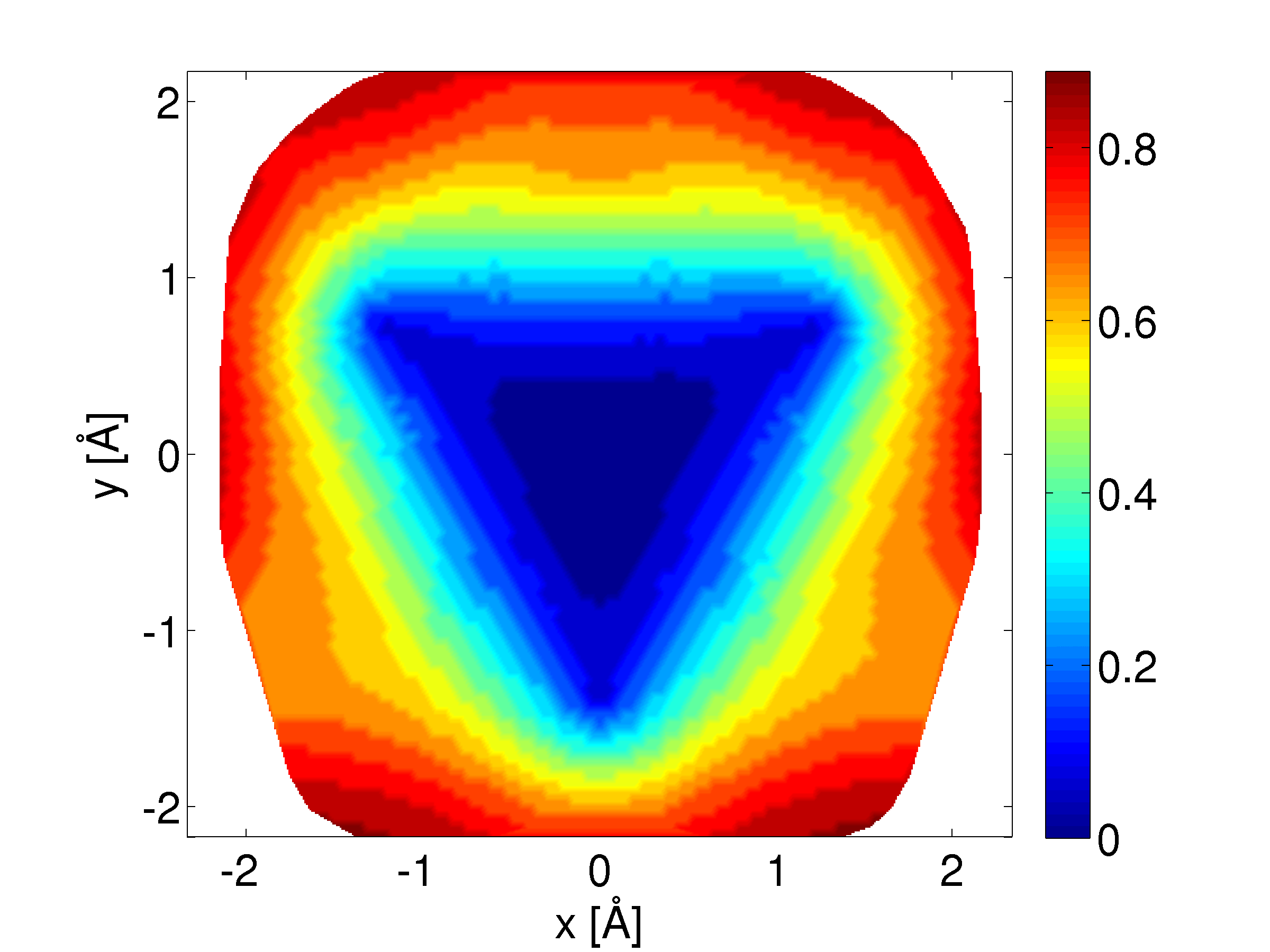}
                \caption{$d_s=\unit[1.85]{\AA}$}
                \label{fig:IrGrW_areas_3}
        \end{subfigure}
        \begin{subfigure}[b]{0.4\linewidth}
                \includegraphics[width=\textwidth]{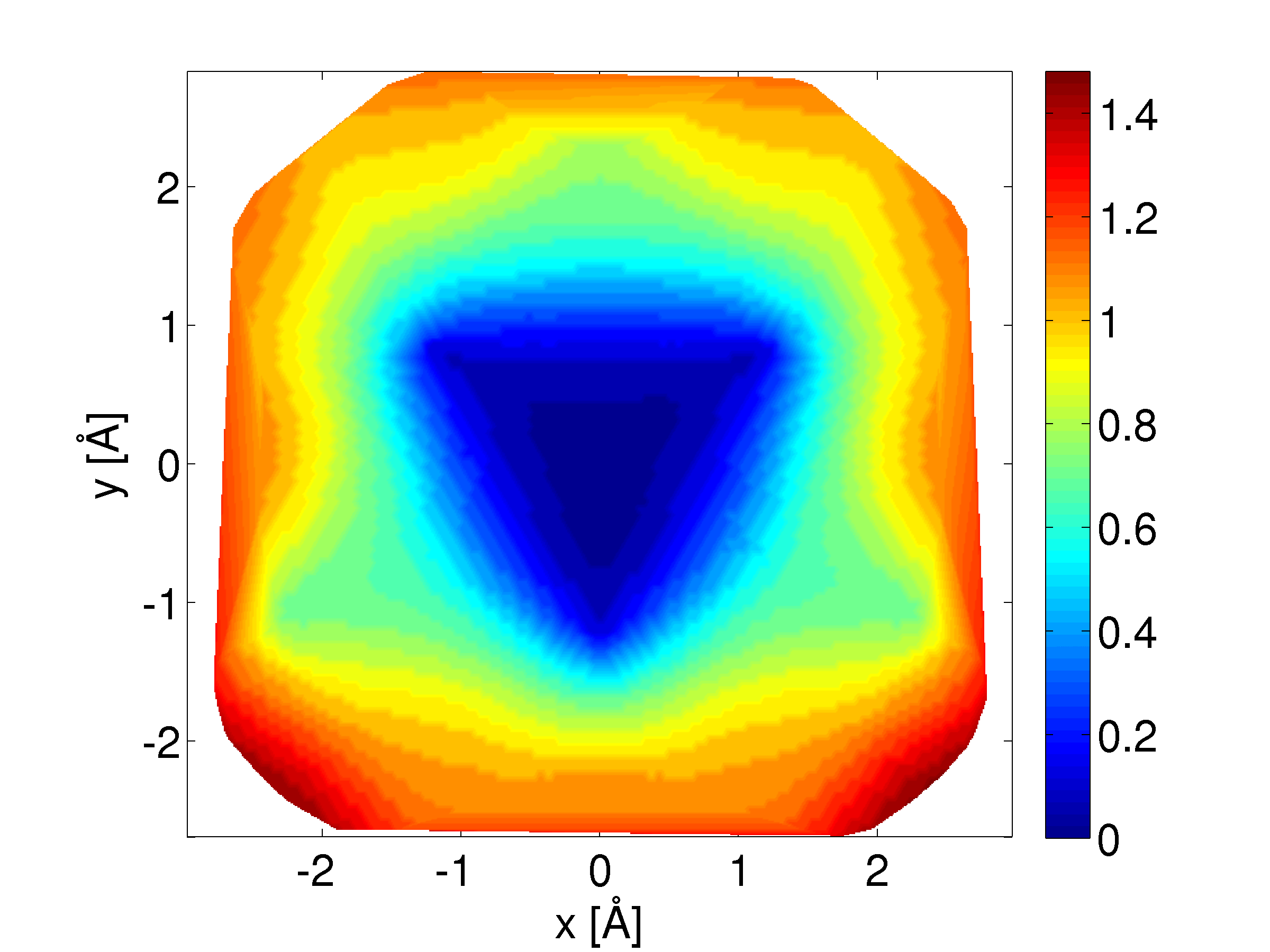}
                \caption{$d_s=\unit[1.30]{\AA}$}
                \label{fig:IrGrW_areas_4}
        \end{subfigure}
   \caption{Contact formed by lowering a tungsten tip onto a graphene/Ir(111) surface. The contact area increases from (a) \unit[6.8]{\AA$^2$}, over (b) \unit[9.1]{\AA$^2$}, and (c) \unit[21.13]{\AA$^2$}, to (d) \unit[35.0]{\AA$^2$}. Different depths of the curved contact areas are coded by color contours with the lowest value set to zero. Please note the different color bars and axes scaling in each panel.}
	\label{fig:IrGrW_areas}
\end{figure*}

Figure~\ref{fig:IrGrW_areas} shows the geometrical shape of the
non-vanishing real contact area for four distances. The different colors denote different depths, ranging from $\sim\unit[0.3]{\AA}$ in figure~\ref{fig:IrGrW_areas_1} to $\sim\unit[1.4]{\AA}$ in figure~\ref{fig:IrGrW_areas_2}. Initially, for larger distances, the shape is rather flat and is dominated by the threefold symmetry of the graphene layer (figure~\ref{fig:IrGrW_areas_1} and~\ref{fig:IrGrW_areas_2}). As the graphene layer gets depressed towards the iridium substrate, the contact area is beginning to show a pronounced bowl shape (figure~\ref{fig:IrGrW_areas_3}), which increases in depth for decreased distance (figure~\ref{fig:IrGrW_areas_4}). Note that for the closest distance (figure~\ref{fig:IrGrW_areas_4}, $d_s=\unit[1.30]{\AA}$) not only the threefold symmetry of the graphene layer is visible in the center, but the edges of the bowl have the fourfold symmetry of the second tip layer.

\begin{figure}[htbp]
\centering
\includegraphics[angle=0, width=1.0\linewidth]{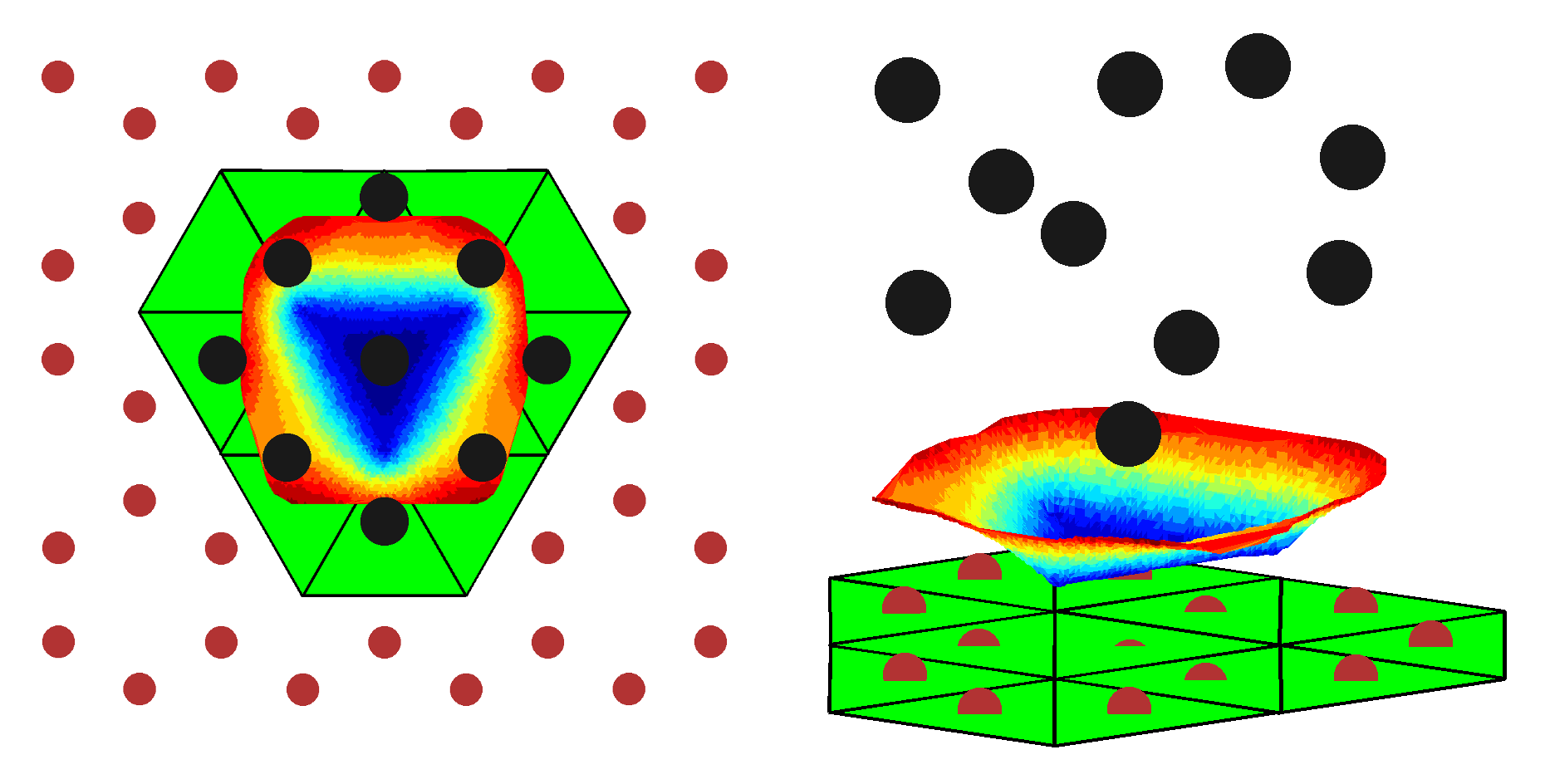}
\caption{Top view (left) and side view (right) of the real contact area resulting from our \textit{ab-initio} approach using Bader atoms (color code depending on height) for a distance of $d_s=\unit[1.85]{\AA}$ (see figure~\ref{fig:IrGrW_areas_3}), compared to the flat contact area from the model by Mo, Turner, and Szlufarska (green)~\cite{mo:09},  Carbon and tungsten atoms are sketched as red and black dots, respectively.}
 \label{fig:IrGrW_A_comp}
\end{figure}

We can also analyze how many carbon atoms are in contact with the tip for each distance and compare the contact area predicted by our approach with the results by Mo, Turner, and Szlufarska~\cite{mo:09}. To this end we count every surface Bader atom that touches our tip as contacting, a different approach than described in the introduction, since we have no pairwise forces at our disposal. In our case, with a cross section of the simulation cell $A_\mathrm{C}\sim\unit[262]{\AA^2}$, and 200 carbon atoms in the graphene layer the contribution per atom to the real contact area is $A_\mathrm{at}=A_\mathrm{C}/200=\unit[1.31]{\AA^2}$. Each carbon atom has 3 nearest neighbors in $d_\mathrm{nn}=\unit[1.42]{\AA}$, 6 next nearest neighbors at \unit[2.46]{\AA}, 3 third nearest neighbors at \unit[2.84]{\AA}, and 6 fourth nearest neighbors at \unit[3.76]{\AA} distance. Already directly after the jump to contact at $d_s=\unit[3.53]{\AA}$, more than one carbon atom is in contact with the tip, although the majority of the contact is formed by the central carbon atom which is responsible for \unit[5.90]{\AA$^2$} of the total \unit[6.82]{\AA$^2$}. This is about 30\% more than the contact area predicted in reference~\citenum{mo:09}, with $4A_\mathrm{at}=\unit[5.25]{\AA^2}$. The area contributed by the central carbon atom alone exceeds the value of $4A_\mathrm{at}$ by $\sim 12\%$. The next nearest and third nearest neighbors begin to play a role at $d_s=\unit[2.18]{\AA}$, contributing to about 7\% of the total area of \unit[16.89]{\AA$^2$}. Here the method by Mo, Turner, and Szlufarska gives  a very comparable area of $13A_\mathrm{at}=\unit[17.05]{\AA^2}$. However, the 9 outermost atoms that contribute 7\% to the contact area in our approach are responsible for nearly 70\% of the contact area in reference~\citenum{mo:09} considering all contacting atoms equally. The situation at $d_s=\unit[1.85]{\AA}$ is visualized in figure \ref{fig:IrGrW_A_comp}, with the method of reference~\citenum{mo:09} still giving an area of $13A_\mathrm{at}=\unit[17.05]{\AA^2}$, while our approach yields \unit[21.13]{\AA$^2$}. Only for the two closest positions at $d_s=\unit[1.51]{\AA}$ and $d_s=\unit[1.30]{\AA}$ more than 13 atoms are in contact, according to the Bader partitioning, and the central 13 are still responsible for 98\% and 87\% of the contact area, respectively. Including the 6 fourth nearest neighbors into the model by Mo, Turner, and Szlufarska~\cite{mo:09}, leads to $19A_\mathrm{at}=\unit[24.92]{\AA^2}$ for both of this distances, while our approach gives $A(1.51)=\unit[28.89]{\AA^2}$ and $A(1.30)=\unit[35.03]{\AA^2}$. Thus, the results are comparable, but the outermost atoms are again over represented compared to our approach. Our model offers higher resolution of the real contact area and allows for a distance dependent contribution of each atom. It is important to note that our contact areas are curved and have a more or less pronounced bowl shape while Mo, Turner, and Szlufarska consider flat contact areas (see figure~\ref{fig:IrGrW_A_comp}). Overall both methods show fair agreement.

Our chosen system, which has been proven to accurately model the interaction between a tungsten tip and moir\'e graphene on Ir(111)~\cite{voloshina:13}, limits our investigation to the attractive region ($d_r\geq\unit[2.24]{\AA}$) and small positive loads ($\unit[2.18]{\AA}\leq d_r<\unit[2.24]{\AA}$). For $d_r<\unit[2.18]{\AA}$, the tip forms bonds with the iridium substrate leading again to negative values of the load. Thus, it is difficult to predict the behavior of the real contact area $A$ dependent on the load $L$. However, we can assume that the interaction potential $E(d_r)$ can be approximated by a Morse potential in the vicinity of the minimum~\cite{morse:29},
\begin{equation}
E_{\mathrm{M}}\left(  d_r\right)=E_{0}\left\{  \left[  1-e^{-\gamma\left(
d_r-d_{0}\right)  }\right]  ^{2}-1\right\}  \quad , 
\label{eq:morse-fct}
\end{equation}
where $E_0=\unit[2.33]{eV}$ is the depth of the potential at the equilibrium position $d_0=\unit[2.24]{\AA}$, which we can get directly from our data. Thus only $\gamma$ has to be fitted, resulting in $\gamma=\unit[4.11]{\AA^{-1}}$. We can now derive the load $L=-\partial E_{\mathrm{M}} / \partial d_r$ yielding
\begin{equation}
L\left(d_r\right)=-2\gamma E_{0}\left[ 1-e^{-\gamma\left(d_r-													d_{0}\right)}\right]e^{-\gamma\left(d_r-d_{0}\right)} \quad .
\end{equation}
Solving this for $d_r$ produces
\begin{equation}
d_r=d_{0}-\dfrac{1}{\gamma}\ln\xi\left(  L\right)  \quad ,
\label{eq:morse-dist}%
\end{equation}
where the dimensionless function $\xi(L)$ is
\begin{equation}
\xi\left(  L\right)  =\dfrac{1\pm\sqrt{1-4u}}{2}~,\qquad\mathrm{with}\qquad
u=\dfrac{L}{2\gamma E_{0}} \quad .
\label{eq:morse-prm}
\end{equation}
Now it is possible to express the real contact area dependent on load using equations~(\ref{eq:Ar}) and~(\ref{eq:morse-dist}), which provides a power law,
\begin{equation}
A\left(  L\right) =A_{0}e^{-\lambda_r\left( d_{0} - \Delta_r\right)} \,\left[  \xi\left(
L\right)  \right]  ^{\frac{\lambda_r}{\gamma}} \quad .
\label{eq:morse-area}
\end{equation}
As $\gamma=\unit[4.11]{\AA^{-1}}$ and $\lambda_r=\unit[4.19]{\AA^{-1}}$ the exponent is very close to 1, thus we arrive at a linear dependence of $A$ on $\xi(L)$, namely $A(L)=C \xi(L)$ with $C=\unit[24.15]{\AA^2}$, which corresponds to an increase of $A$ with $L$ to the power of $\frac{1}{2}$.

While we believe that our \textit{ab-initio} approach using the QTAIM for calculating the real contact area is intuitive and accurate, its limitations have to be discussed as well. As the charge density is required to perform the Bader partitioning our approach is limited to system sizes where \textit{ab-initio} calculations are still feasible. This limits us to a single asperity case at the moment, where only a handful of atoms interact with the surface. However, since our system was used to explain some experimental results~\cite{voloshina:13,garhofer:phd}, we are confident that our scheme is applicable to real systems, albeit only for sharp tips and low loads. With the ever increasing power of modern computers and better scaling codes, on the other hand, it might soon be feasible to calculate systems with thousands or millions of atoms and hence to study interesting phenomena such as multi-asperity contacts. Thus, our approach might be used in MD calculations as well to directly investigate the influence of the real contact area on frictional forces~\cite{vernes:12}.

\section{Conclusion}
\label{sec:conclusion}

We propose a new {\it ab-initio} approach for calculating the real
contact area between a tip and a surface. We apply Bader's quantum theory of atoms in molecules (QTAIM) to determine the volumes and shapes of the atoms in contact together with their contact areas at each given distance. We define a specific density cutoff $\rho_{\mathrm{cut}}$ for this partitioning to confine the Bader volumes to realistic values. This cutoff density is obtained by using the discontinuity in the $d_r$ versus $d_s$ curve to define the initial point of contact in a perspicuous way, which in the examined system occurs due to a jump to contact, commonly observed in AFM experiments. This defines a lower bound for the cutoff density which is then the optimal value, since $\rho_{\mathrm{cut}}$ needs to be minimized to include the maximum number of electrons. Thus, our approach remains essentially \textit{ab-initio}, as the only parameter needed can be determined from properties of the system. We believe that the jump to contact is a less ambiguous way to define the onset of contact than using a partitioning of the interaction in long- and short-range regions~\cite{mo:09}, or equating contact with repulsive interactions~\cite{burnham:91,cheng:10}.

For decreasing the real tip-sample distance $d_r$ an exponential increase of the real contact area $A$ is found. This is a combined effect of the jump to
contact and the preferred distance of the tip apex relative to the
surface atom below it, which first jumps up to meet the tip and then
gets pressed below its equilibrium position for closer separations. As $d_s$ is linear dependent on $d_r$, we can also express the exponential relation $A(d_r)$ through $d_s$, which can be better controlled in experiments than $d_r$.

\section*{Acknowledgments}
The authors would like to thank A. Garhofer for providing some of the
relaxed structures and helping with the selection of computational
parameters, as well as F. Mittendorfer for fruitful discussions.  This
work was funded by the ``Austrian COMET-Program'' (project XTribology,
no. 824187) via the Austrian Research Promotion Agency (FFG) and the
Province of Nieder\"osterreich, Vorarlberg and Wien and has been
carried out within the ``Excellence Centre of Tribology'' (AC2T
research GmbH) and at Vienna University of Technology. P.M., J.R., and
G.F. acknowledge the support by the Austrian Science Fund (FWF) [SFB
  ViCoM F4109-N13]. This work was supported in part also by COST
Action MP1303. The authors also appreciate the ample support of
computer resources by the Vienna Scientific Cluster
(VSC). Fig.~\ref{fig:IrGrW_unitCell} in this paper was created with
the help of the VESTA code~\cite{vesta:11}. 
\bibliography{../../../PhD_Papers/Bib}
\end{document}